\documentclass[runningheads]{llncs}
\usepackage{graphicx}
\usepackage{amsmath}
\usepackage{amsfonts}
\usepackage{multirow}
\usepackage{hyperref}

\begin{document}

\title{Denoising of 3D MR images using a voxel-wise hybrid residual MLP-CNN model to improve small lesion diagnostic confidence}

\titlerunning{Denoising 3D MRI with small lesion}

\author{Haibo Yang\inst{1,2} \and Shengjie Zhang\inst{1,2} \and Xiaoyang Han\inst{1,2} \and Botao Zhao\inst{1,2} \and Yan Ren\inst{3}\and Yaru Sheng\inst{3} \and Xiao-Yong Zhang\inst{1,2}}


\authorrunning{H. Yang et al.}

\institute{Institute of Science and Technology for Brain-Inspired Intelligence, Fudan University, Shanghai, China.
\and
Key Laboratory of Computational Neuroscience and Brain-Inspired Intelligence (Fudan University), Ministry of Education, China
\\ \email{xiaoyong\_zhang@fudan.edu.cn}
\and Department of Radiology, Huashan Hospital, Fudan University, Shanghai, China}

\maketitle  

\begin{abstract}
Small lesions in magnetic resonance imaging (MRI) images are crucial for clinical diagnosis of many kinds of diseases. However, the MRI quality can be easily degraded by various noise, which can greatly affect the accuracy of diagnosis of small lesion. Although some methods for denoising MR images have been proposed, task-specific denoising methods for improving the diagnosis confidence of small lesions are lacking. In this work, we propose a voxel-wise hybrid residual MLP-CNN model to denoise three-dimensional (3D) MR images with small lesions. We combine basic deep learning architecture, MLP and CNN, to obtain an appropriate inherent bias for the image denoising and integrate each output layers in MLP and CNN by adding residual connections to leverage long-range information. We evaluate the proposed method on 720 T2-FLAIR brain images with small lesions at different noise levels. The results show the superiority of our method in both quantitative and visual evaluations on testing dataset compared to state-of-the-art methods. Moreover, two experienced radiologists agreed that at moderate and high noise levels, our method outperforms other methods in terms of recovery of small lesions and overall image denoising quality. The implementation of our method is available at \url{https://github.com/laowangbobo/Residual\_MLP\_CNN\_Mixer}.

\end{abstract}

\section{Introduction} 
Magnetic resonance imaging (MRI) is a high-resolution and non-invasive medical imaging technique that plays an important role in clinical diagnosis\cite{jiang2019novel,mohan2014survey,zhang2015denoising}. In particular, the diagnosis of small lesions in vivo, such as white matter hyperintensities (WMH) lesions, cerebral small vessel disease (CSVD), stroke, etc, depends largely on image quality \cite{gonzalez2012clinical,ovbiagele2006cerebral,zwanenburg2017targeting}. However, the MRI quality can be often degraded by the noise generated during image acquisition\cite{gudbjartsson1995rician}, thereby affecting detection, segmentation\cite{caligiuri2015automatic,wu2019skip} and diagnosis accuracy of small lesions \cite{jiang2017seizure,khademi2011robust}.

Previous non-learning methods for MRI denoising can be divided into filtering-based and transform-based methods. Filtering-based methods apply linear or non-linear filters to remove noise in MR images, such as non-local means (NLM) filter\cite{coupe2006fast}, Perona–Malik (PM) model\cite{perona1990scale}, and anisotropic diffusion filter (ADF)\cite{sijbers1999adaptive}. Transform-based methods reduce noise with a new image representation by transforming images from spatial domain into other domains, including wavelet-based methods\cite{anand2008mri} and discrete cosine transform -based methods\cite{hu2012improved}. These traditional non-learning methods suffer from several limitations, such as relying on complex algorithms based on image priors, heavy computational burden, etc.

Currently, deep learning (DL) methods have been introduced into MRI denoising and achieved better performance than traditional methods\cite{ledig2017photo,zhang2017beyond}. For example, several pioneering studies have been reported. Jiang et al\cite{jiang2018denoising} and You et al\cite{you2019denoising} applied a plain convolutional neural network (CNN) to learn the noise in MR images. Ran et al\cite{ran2019denoising} proposed a residual encoder–decoder Wasserstein generative adversarial network (RED-WGAN) to directly generate the noise-free MR images. Although the results produced by these DL methods are promising, their performance are not satisfied for denoising MR images. On the other hand, these methods have not been evaluated in clinical diagnosis. Thus, novel models are still needed to improve MRI denoising quality, especially for small lesion diagnosis that is easily contaminated by image noise.

In this paper, we propose a new DL method based on a voxel-wise hybrid residual MLP-CNN model to denoise three-dimensional (3D) MR images. We combine multilayer perceptrons (MLPs) and convolutional neural networks (CNNs) to obtain appropriate intrinsic bias for image denoising, thereby providing an appropriate number of parameters to avoid overfitting and thus improve model performance. Additionally, we integrate each output layers in MLP and CNN by adding residual connections to leverage long-range information. The structure of residual MLP is inspired by masked autoencoders (MAE)\cite{he2021masked}, which is used to reconstruct the random missing pixels. MAE uses vision transformer (ViT)\cite{dosovitskiy2020image} as encoder to reconstruct random masked patches in a image. Due to computational limitation, instead of using ViT as encoder, a simple MLP with residual connection was used to encode each noising voxel in MR images in our model. In short, the main contributions of this paper are: (a) our method can significantly denoise 3D MR images; (b) our method shows a superior performance for the recovery of small lesions in MR images compared to several state-of-the-art (SOTA) methods; (c) the diagnosis confidence of small lesion is confirmed by experienced radiologists.

\section{Methods}
The main purpose of MRI denoising is to recover the clean image $X \in R^{H \times W}$ from noisy image $Y \in R^{H \times W}$. Assuming that $\hat{Y}=f(Y)$ is the denoised image by $f$, MR image denoising can be simplified to find the optimal approximation of $f$:

\begin{equation}
\arg \min _{f}\|X-\hat{Y}\|_{2}^{2}
\end{equation}

\begin{figure}
\includegraphics[width=\textwidth]{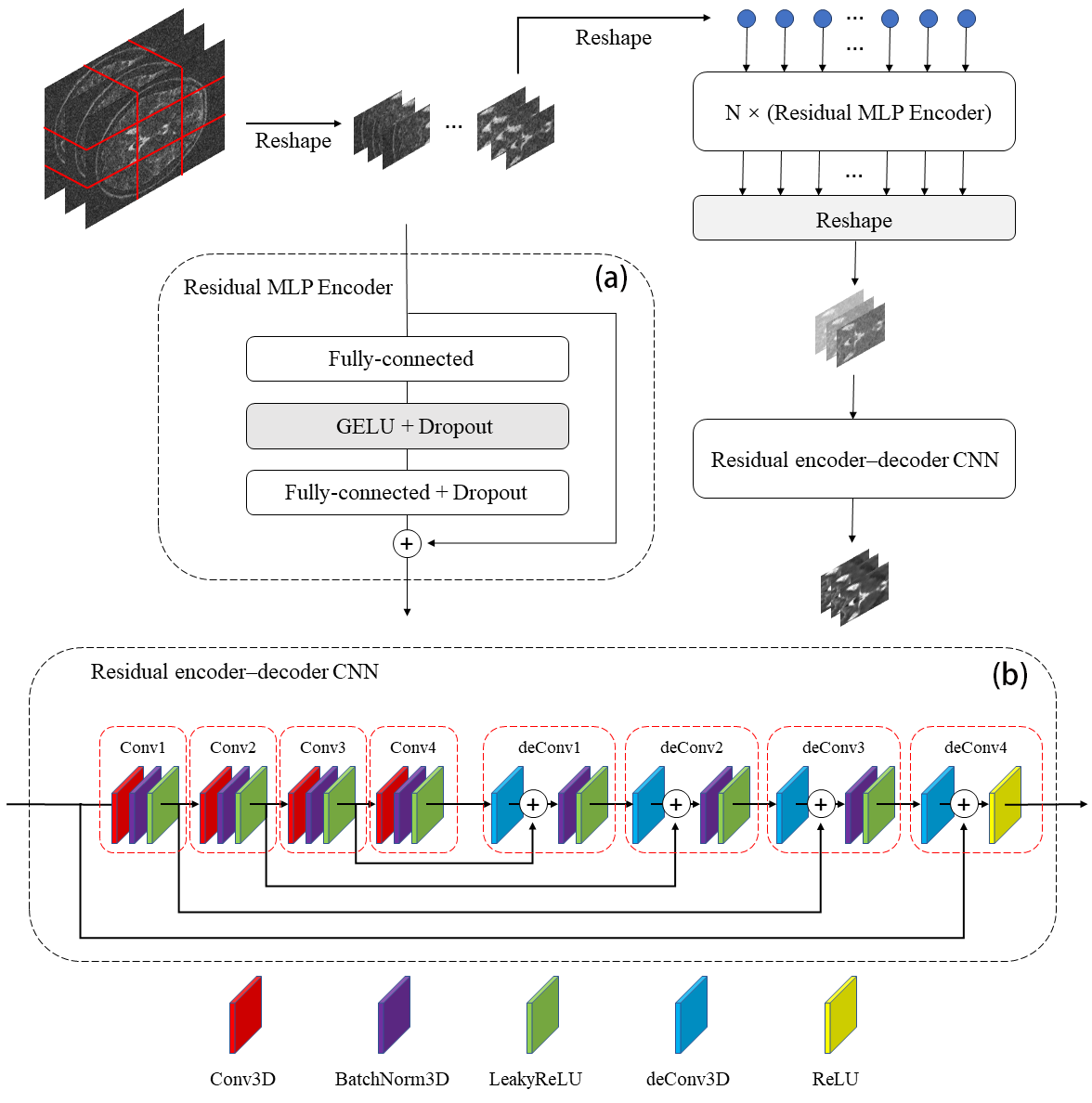}
\caption{Overall architecture of our proposed model. It consists of several residual MLP encoders and a residual encoder-decoder CNN. The details of residual MLP encoders and residual encoder-decoder CNN are shown in two black dotted boxes (a) and (b) respectively.} \label{network}
\end{figure}

An overview of our architecture is illustrated in Fig.~\ref{network}. It consists of several residual MLP encoders and a residual encoder-decoder CNN. To process the 3D images, we reshape the noisy image $\mathbf{X} \in \mathbb{R}^{ H \times W \times C}$ into a sequence of flatten 3D patches, $\mathbf{X}_{p} \in \mathbb{R}^{K \times\left(P^{2} \cdot C\right)}$, where $(H, W)$ is the resolution of the original image, $C$ is the number of slices, $(P, P)$ is the resolution of each image patch, $K$ is the resulting number of patches.

Each MLP block contains two fully-connected layers and a nonlinearity applied independently to each flatten patches. Residual MLP encoders can be written as follows:

\begin{equation}
\mathbf{z}_{l}=\mathbf{z}_{l-1}+\mathbf{W}_{2} \delta\left(\mathbf{W}_{1} \text { LN }(\mathbf{z}_{l-1})\right), \quad \text { for } l=1 \ldots L
\end{equation}

Here $\delta$ is an element-wise nonlinearity (GELU). $\mathbf{z}_{l}$ is result of the $lth$ MLP encoder, $\mathbf{z}_{0}$ serves as one flatten 3D patches, the size of $\mathbf{z}_{0}$ is  $(P^{2} \cdot C \times 1 \times 1)$. $\mathbf{W}_{1}$ and $\mathbf{W}_{2}$ is the first and the second fully-connected layer weights. L is the number of Residual MLP encoders. Layernorm (LN) is applied before every block, and residual connections after every block.

\begin{table}
\centering
\caption{Qualitative Image Evaluation Criteria and Scoring Definitions.}\label{standard tab}
\begin{tabular}{l|l|l}
\hline
\bfseries Score     & \bfseries Small Lesion Conspicuity                                & \bfseries Overall Image Quality                                                                                                                                              \\\hline
1$\sim$2  & No visualization                                        & No noise reduction                                                                                                                                                 \\\hline
3$\sim$4  & Poor visualization                                      & \begin{tabular}[c]{@{}l@{}}Noise was reduced slightly, \\ and only a few tissue contours\\ were discernible\end{tabular}                                             \\\hline
5$\sim$6  & \begin{tabular}[c]{@{}l@{}}Visualized with location \\but have blurred margins  \end{tabular}      & \begin{tabular}[c]{@{}l@{}}Some noise is reduced \\and some details of the tissue \\ could be identified, but not clear\end{tabular}                                 \\\hline
7$\sim$8  &\begin{tabular}[c]{@{}l@{}} Most   of lesions are clearly visible, \\some are blurred \end{tabular}& \begin{tabular}[c]{@{}l@{}}Noise is almost reduced, \\tissue details can be clearly identified, \\ some noise remains \\but does not affect the diagnosis\end{tabular} \\\hline
9$\sim$10 & \begin{tabular}[c]{@{}l@{}}Almost   the same as \\the noise-free image   \end{tabular}            & Remove the noise completely \\ \hline                                                                                                                                     
\end{tabular}
\end{table}

Residual encoder-decoder CNN has an encoder–decoder structure composed of $J$ convolutional and $J$ deconvolutional layers. Skip residual connections link the
corresponding convolution-deconvolutional layer pairs. Except for the last layer, the rest layers contain a 3D convolution, a batch-normalization and a LeakyReLU operation in order, while the last layer only contains a 3D convolution and a ReLU operation. Residual encoder-decoder CNN can be written as follows:

\begin{align}
\mathbf{x}_{j}&=\sigma\left(\text { BN }(\mathbf{H}_{j}\mathbf{x}_{j-1})\right), &\quad \text { for } j=1 \ldots J \\
\mathbf{y}_{j}&=\sigma\left(\text { BN }(\mathbf{U}_{j}\mathbf{y}_{j-1}+\mathbf{x}_{J-l})\right), &\quad \text { for } j=1 \ldots J \\
\mathbf{y}_{J}&=\theta(\mathbf{U}_{J}\mathbf{y}_{J-1}+\mathbf{x}_{0}), \\
\mathbf{y}_{0}&=\mathbf{x}_{J},
\end{align}

Here $\sigma$ is a LeakyReLU operation, $\theta$ is a LeakyReLU operation. $\mathbf{x}_{j}$ is result of the $jth$ convolution layer, $\mathbf{y}_{j}$ is result of the $jth$ deconvolution layer. $\mathbf{H}_{j}$ is the convolutional weight, $\mathbf{U}_{j}$ is the deconvolutional weight. $\mathbf{x}_{0}$ is reshaped from $\mathbf{z}_{L}$, The size of $\mathbf{x}_{0}$ is changed from $(P^{2} \cdot C \times 1 \times 1)$ to $(P \times P \times C)$. $BN$ represents batch-normalization operation.

\begin{table}
\centering
\caption{Ablation study of models with different block combinations. The average PSNR and SSIM are measured on validation set with 15$\%$ noise level. MLP stands for the residual MLP encoders block and CNN stands for the residual encoder-decoder CNN block. }\label{Ablation}
\begin{tabular}{lcc}
\hline
Model   & PSNR             & SSIM            \\ \hline
MLP+MLP      & 29.8472          & 0.8118          \\
CNN+CNN      & 29.8285          & 0.8131          \\
MLP+CNN      & \textbf{32.2679} & \textbf{0.8690} \\ \hline
\end{tabular}
\end{table}

Mean squared error (MSE) loss was used to train the propose model. It can be calculated as follows:

\begin{equation}
L_{MSE}=\frac{1}{HWC}\|G(X)-Y\|^{2}
\end{equation}

Where $H$, $W$ and $C$ represent the dimensions of the image, $X$ is the noisy image, $Y$ is the noise-free image, $G$ represents the function of the model. Note that the concise DL-based architecture and MSE loss make our model easy to train.

\begin{table}
\centering
\caption{The average PSNR and SSIM measures of different methods on testing images with different noise levels.}\label{UKB tab}
\begin{tabular}{lcccccccc}
\hline
{\multirow{2}{*}{Method}} & \multicolumn{2}{c}{3\%}            &           & \multicolumn{2}{c}{9\%}            &           & \multicolumn{2}{c}{15\%}           \\ \cline{2-9} 
\multicolumn{1}{c}{}                        & PSNR             & SSIM            &           & PSNR             & SSIM            &           & PSNR             & SSIM            \\ \hline
BM4D                                        & 31.2357          & 0.5184          &           & 22.5407          & 0.3590          &           & 18.1125          & 0.2915          \\
PRINLM                                      & 37.0051          & 0.9167          &           & 29.7914          & 0.6699          &           & 24.6919          & 0.3994          \\
RED                                         & 35.7440          & 0.8845          &           & 31.4146          & 0.8363          &           & 28.8303          & 0.7721          \\
RED-WGAN                                    & 35.9356          & 0.9006          &           & 32.0610          & 0.8572          &           & 29.5153          & 0.8011          \\
Ours                                        & \textbf{38.8119} & \textbf{0.9378} & \textbf{} & \textbf{34.8806} & \textbf{0.9000} & \textbf{} & \textbf{32.4347} & \textbf{0.8536} \\ \hline
\end{tabular}
\end{table}

\section{Experiments and Discussion}

\subsection{Dataset}
The MRI data from UK Biobank (\url{http://www.ukbiobank.ac.uk}) were used to validate the performance of the proposed method. Details of the image acquisition and processing are freely-available on the website (\url{http://biobank.ctsu.ox.ac.uk/crystal/refer.cgi?id=2367}).

120 T2-FLAIR brain MRI volumes (720 images) with white matter hyperintensities (WMH) lesions were randomly selected. 100 brain volumes (600 images) were used as the training set, 10 volumes (60 images) used as the validation set, the rest 10 volumes (60 images) used as testing set. For each volume, the central slice and its neighboring slices were extracted to form one input 3D MRI volume. The number of slices that we use to denoise 3D image volume is 6. All the image volumes were reshaped into size of $176 \times 256 \times 6$. The training MRI volumes were cropped into $16 \times 16 \times 6$ patches with the stride 10, which means a total of 42,500 training patches with size of $16 \times 16 \times 6$ were acquired.

\begin{figure}
\includegraphics[width=\textwidth]{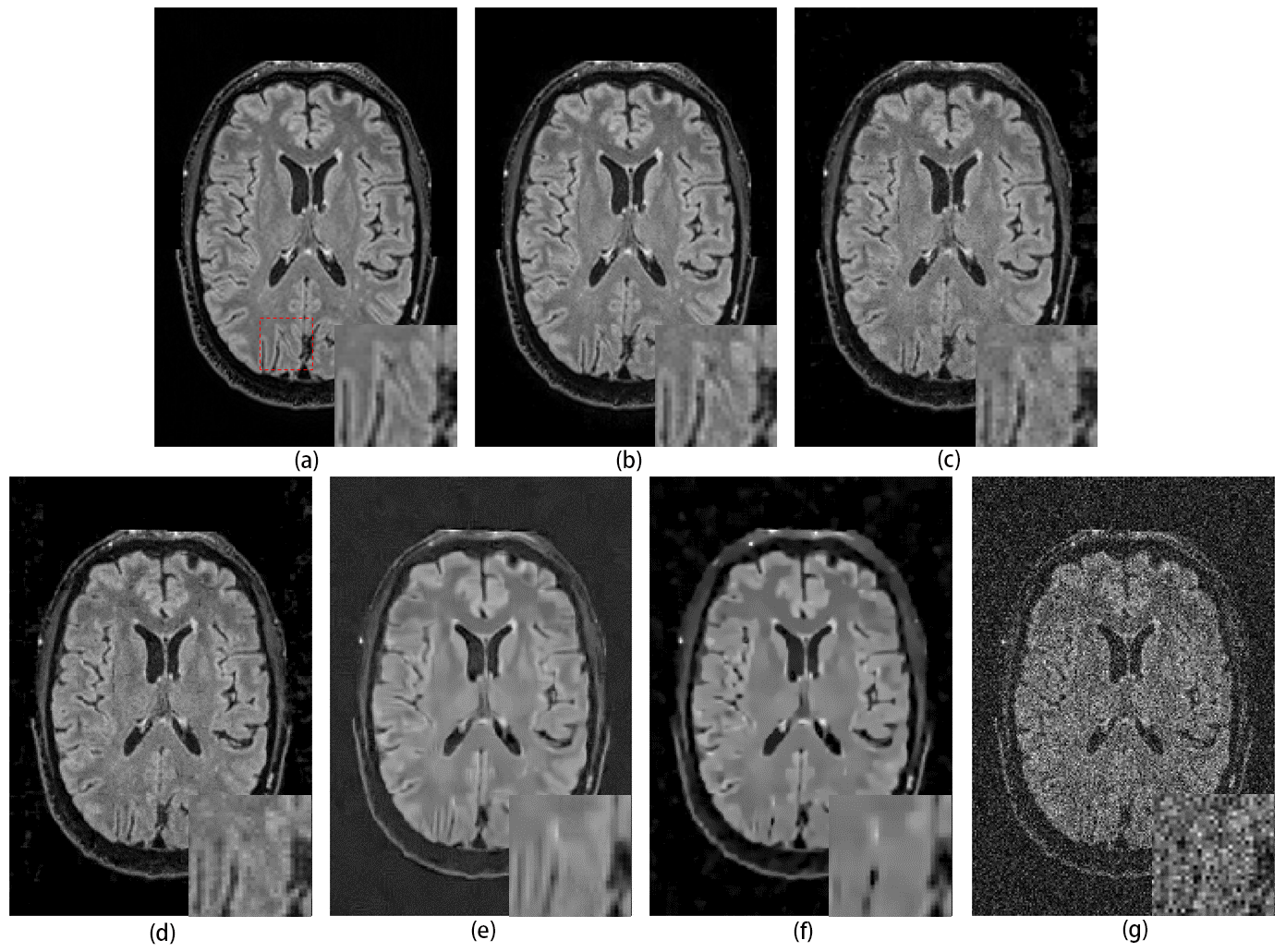}
\caption{A visualization of different denoising methods on T2-FLAIR images with 15$\%$ Rician noise. (a) Noise-free image, (b) Our method, (c) RED-WGAN, (d) RED, (e) BM4D, (f) PRINLM, (g) Noisy image. The small picture in the lower right corner of each image is an amplified version of the area inside the red box in (a).} \label{issue_denoise}
\end{figure}

Previous studies have shown that the noise in MRI is governed by a Rician distribution, both real and imaginary parts are corrupted by Gaussian noise with equal variance\cite{li2020mri}. We simulated noisy images by manually adding Rician noise to the images, as follows: 

\begin{equation}
A=\sqrt{\left(I+\alpha L n_{r}\right)^{2}+\alpha L n_{i}^{2}}
\end{equation}

Here $I$ is the original 3D image volume, $\alpha$ is the maximum value of $I$, $L$ is the noise level, $n_{r}$ and $n_{i}$ are independent and identical $\mathcal{N}(0,1)$ Gaussian distributed noise, and $A$ is the noisy image. Three levels of noise, 3$\%$, 9$\%$ and 15$\%$ were added into original T2-FLAIR volumes.

\subsection{Training Details}
In our experiments, the kernel size of 3D convolution and deconvolution operation is $3 \times 3 \times 3$. The Adam\cite{kingma2014adam} algorithm was used to optimize the proposed model. The initial learning rate is $5e-4$. We utilize Pytorch as computing framework to implement the proposed method. The network is trained on two NVIDIA RTX A6000 GPUs with 48 GB memory.

\begin{figure}
\includegraphics[width=\textwidth]{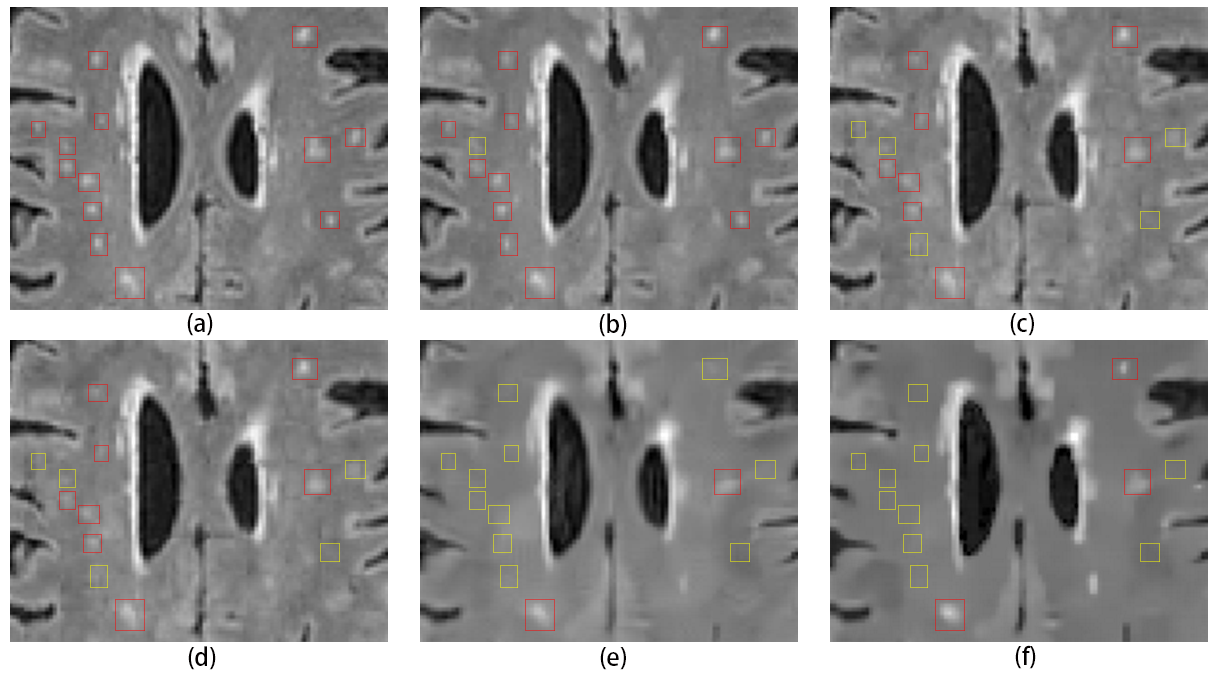}
\caption{One zoomed denoised example with WMH lesions from the testing set with 15$\%$ Rician noise. (a) Noise-free image, (b) Our method, (c) RED-WGAN, (d) RED, (e) BM4D, (f) PRINLM. The area in the box is WMH lesion labeled by experienced reader, the red box indicates that WMH lesions can be identified from the image, while the yellow box indicates that WMH lesion can not be identified from the image. } \label{lesion}
\end{figure}

\subsection{Evaluation Methods}
To validate the performance of the proposed method, two quantitative metrics were employed, including peak signal-to-noise ratio (PSNR) and structural similarity index measure (SSIM)\cite{kala2018adaptive}. PSNR calculates the distortion between stored images and ground truth images. The metrics of SSIM is calculated based on three comparative measurements, including luminance, contrast, and structure.

To further demonstrate the diagnostic confidence of our method, we invited two experienced radiologists as readers to score the overall image quality and the small lesion conspicuity in the restored testing images. The scoring criteria is shown in Table~\ref{standard tab}. 

\subsection{Results and Discussion}
Ablation experiments conducted at validation set with 15$\%$ noise level show the effectiveness of combination of residual MLP encoders and residual encoder-decoder CNN. Table~\ref{Ablation} shows the performance improvements when we combine the MLP block and the CNN block.

\begin{table}
\centering
\caption{Summary of Image Evaluation.}\label{rate tab}
\begin{tabular}{l|l|cc|ccc|ccc}
\hline
\multirow{2}{*}{Reader}  & \multirow{2}{*}{Method} & \multicolumn{2}{c|}{3\%}                                   & \multicolumn{1}{l}{} & \multicolumn{2}{c|}{9\%}                                   & \multicolumn{1}{l}{} & \multicolumn{2}{c}{15\%}                                  \\ \cline{3-10} 
                         &                         & \multicolumn{1}{l}{Lesions} & \multicolumn{1}{l|}{Overall} & \multicolumn{1}{l}{} & \multicolumn{1}{l}{Lesions} & \multicolumn{1}{l|}{Overall} & \multicolumn{1}{l}{} & \multicolumn{1}{l}{Lesions} & \multicolumn{1}{l}{Overall} \\ \hline
\multirow{5}{*}{Reader1} & BM4D                    & 7.4                         & 8.3                          &                      & 5.8                         & 6.3                          &                      & 4.7                         & 5.2                         \\
                         & PRINLM                  & 8.2                         & 8.7                          &                      & 6.7                         & 6.8                          &                      & 5.8                         & 6.3                         \\
                         & RED                     & 9.1                         & 9.0                          &                      & 8.0                         & 8.2                          &                      & 7.2                         & 7.5                         \\
                         & RED-WGAN                & 8.8                         & 8.8                          &                      & 8.0                         & 8.2                          &                      & 7.7                         & 7.5                         \\
                         & Ours                    & \textbf{9.6}                & \textbf{9.6}                 & \textbf{}            & \textbf{8.5}                & \textbf{8.5}                 & \textbf{}            & \textbf{8.4}                & \textbf{8.2}                \\ \hline
\multirow{5}{*}{Reader2} & BM4D                    & 8.8                         & 8.9                          &                      & 6.2                         & 6.2                          &                      & 5.6                         & 5.6                         \\
                         & PRINLM                  & \textbf{9.7}                & \textbf{9.7}                 &                      & 6.8                         & 6.8                          &                      & 6.2                         & 6.2                         \\
                         & RED                     & 9.3                         & 9.3                          &                      & 7.2                         & 7.3                          &                      & 6.8                         & 6.8                         \\
                         & RED-WGAN                & 9.1                         & 9.1                          &                      & 7.8                         & 7.8                          &                      & 6.9                         & 7.1                         \\
                         & Ours                    & 9.2                         & 9.2                          &                      & \textbf{8.1}                & \textbf{8.1}                 & \textbf{}            & \textbf{7.6}                & \textbf{7.7}                \\ \hline
\end{tabular}
\end{table}

To demonstrate the performance of the proposed method, four different SOTA methods are used for the comparison, including BM4D\cite{maggioni2012nonlocal}, PRINLM\cite{manjon2010adaptive}, RED\cite{mao2016image}, RED-WGAN\cite{ran2019denoising}.

The average quantitative results from different methods at various noise levels on testing set are summarized in Table~\ref{UKB tab}. The proposed method significantly outperforms other methods on all metrics under various noise levels. At 3$\%$ noise level, the traditional non-learning method, PRINLM, yields a better performance than methods based deep learning, RED and RED-WGAN. While as the noise level increases, at 9$\%$ and 15$\%$ noise level, the deep learning based methods outperform traditional non-learning methods, especially on the SSIM metrics. 

Fig.~\ref{issue_denoise} provides a visual evaluations of different methods for a 15$\%$ noise level task. Compared with other methods, the proposed method is most similar to the noise-free image in terms of the clarity of brain wrinkles and folds. Note that some brain structural details are missing from the RED and RED-WGAN results, while most brain structural details are missing from the BM4D and PRINLM results. 

Fig.~\ref{lesion} shows representative small lesions recovery results by different methods. Our method restores almost all labeled WMH lesions, which is the best of all methods. In contrast, the traditional non-learning methods, such as BM4D and PRINLM, miss most of labeled WMH lesions. Table~\ref{rate tab} shows the scores of the two readers on the testing dataset. Reader1 believes that the proposed method outperforms other methods in both small lesion recovery scores and overall image quality scores at all noise levels, while Reader2 believes that PRINLM outperforms other methods at 3$\%$ noise level. Both readers agree that our proposed method outperforms the other methods at 9$\%$ and 15$\%$ noise levels.

\subsection{Conclusions}
We propose a voxel-wise hybrid residual MLP-CNN model to denoise 3D MR images and improve diagnostic confidence for small lesions. We compared the proposed method with four SOTA methods on 720 T2-FLAIR brain images with WMH lesions by adding 3$\%$, 9$\%$ and 15$\%$ noise. The results demonstrate that our method outperforms other methods in quantitative and visual evaluations at testing dataset. In addition, two experienced readers agreed that our method outperforms other methods in terms of small lesions recovery and overall image denoising quality at moderate and high noise levels. 

\subsubsection{Acknowledgments.} This study was supported in part by the National Natural Science Foundation of China (81873893, 82171903), Science and Technology Commission of Shanghai Municipality (20ZR1407800), and Shanghai Municipal Science and Technology Major Project (No.2018SHZDZX01).

\bibliographystyle{unsrt}
\bibliography{mybibliography}
\end{document}